# High Power and Wideband Terahertz Modulator Using *c*-Axis Current Controlled Layered Superconductors

Alireza Kokabi, Hamed Kamrani, and Mehdi Fardmanesh, *Senior Member, IEEE*

***Abstract*—The possibility of applying stacks of capacitive and inductively coupled and synchronized Josephson junctions as either high radiation power or wideband current controlled terahertz modulator is proposed and investigated. The properties of such modulators, which can be fabricated using the mesa structure of the layered high-$T_c$ superconductors, are obtained numerically and analytically in the presence and absence of the external magnetic field with different vortex lattices. Based on these analyses, the peak frequency deviation of modulated signal with respect to input signal is calculated. Typical spectrum of the radiated signal is also obtained in this approach. In addition, the variation of the peak frequency deviation with respect to carrier frequency for different magnetic configurations is obtained analytically. The results for these calculations show carrier frequencies from subtrahertz to multiterahertz.**

## I. Introduction

Terahertz modulators are the topic of some recent tremendously high frequency (THF) device oriented works due to the technology demands for the utilization of the terahertz band in the materials science, biology, security checking, high-bandwidth communications, and so on. Due to the above needs, structures such as quantum dots, tunable photonic crystals, and metamaterials are considered to be applied as terahertz signal modulators [1-9]. Type-II superconductors that can emit terahertz electromagnetic wave due to the flux-flow effect of Abrikosov vortices are other possible candidates for this purpose [10].

Since the terahertz radiation is observed from the stack of intrinsic Josephson junctions in the mesa structure of layered high-$T_c$ superconductors [11], it seems that they can also be applied in many quasi-optical applications. In these structures, the synchronization of high-frequency oscillations in all Josephson junctions can produce terahertz radiation [12]. Such a condition occurs in the presence of applied magnetic field by the proposed mechanism of the flux-flow radiation and in the absence of the magnetic field due to the ac Josephson effect [13-15]. In both cases, the excitation of the internal cavity resonance that can be followed by a large number of junctions through a chain reaction, results in the synchronization of the junction oscillations [16]. In this mechanism, which is known as Fiske modes, the frequency of such an in-phase resonance is controlled by the width of the junctions. Depending on this geometric parameter, two kinds of oscillatory regimes might occur. In the case of short junctions satisfying $L_x < 3\lambda_c$, we deal with monochromatic standing-wave oscillation at the Josephson frequency. On the other hand, in the system of long junctions with $L_x > 3\lambda_c$, vortex-antivortex excitations lead to solitonic modes and nonlinearity effects whose fundamental harmonic is equal to the half Josephson frequency.

In the case of zero external magnetic field to the best of our knowledge yet, it is not clear why the superconductivity phase differences of a large number of junctions is synchronized and results in the intense radiation. However, the phase kinks are proposed for the mechanism of pumping energy into the plasma oscillations [15].

Based on these phase dynamics, here we investigate the effects of the electrical bias on the radiation frequency of the mesa sample. Neglecting nonlinearities in this relation, we propose the application of the stacked junctions as an extremely wideband frequency modulators both in the presence and absence of the magnetic bias. Such a modulator is controlled by the external current, which can be in a wide range of frequencies. Due to the importance of the linear variation of the instantaneous frequency to the controlling signal, we also investigate this property based on the numerical and analytical models. It should also be noted that the other types of the layered superconductors are previously applied as terahertz detectors with different mechanisms [].

## II. Theoretical Model

The frequency modulation for a baseband signal of $x_m(t)$ that carrying data and should be transmitted at the carrier frequency of $f_c$ is defined as [19]

A. Kokabi is with the Department of Electrical Engineering, Sharif University of Technology, Tehran, Iran, and P.O. Box: 11365-11155 (corresponding author to provide phone: +98-21-66165989; fax: +98-21-66165990; e-mail: kokabi@ee.sharif.ir).

H. Kamrani is with the Department of Electrical Engineering, Sharif University of Technology, Tehran, Iran, and P.O. Box: 11365-11155.

M. Fardmanesh is with the Department of Electrical Engineering, Sharif University of Technology, Tehran, Iran, and P.O. Box: 11365-11155.

$$y(t) = A_c \cos\left(2\pi f_c t + 2\pi f_\Delta \int_0^t x_m(\tau)\, d\tau\right). \quad (1)$$

Here $A_c$ is the carrier's amplitude, and $f_\Delta$ is the peak frequency deviation, which represent the maximum shift away from $f_c$. Equation (1) leads to instantaneous frequency of

$$f = f_c + f_\Delta x_m(t). \quad (2)$$

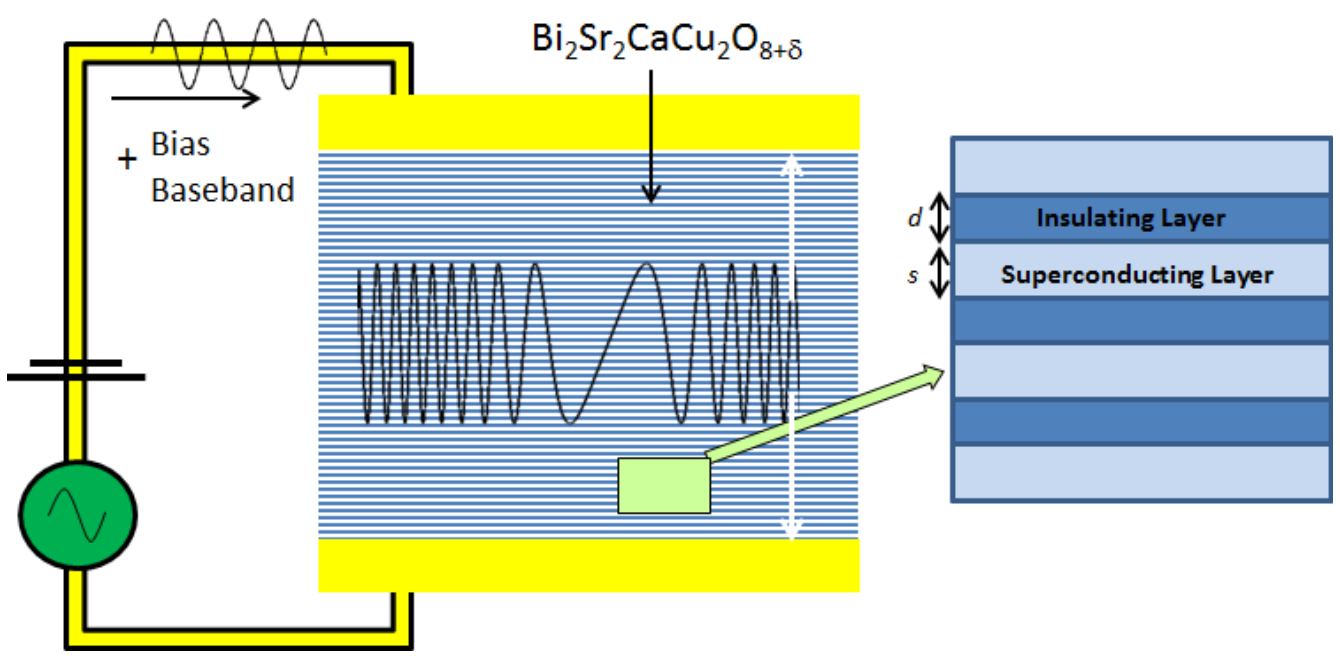

Fig. 1. Bias and baseband signal applied to the stacked junctions

The proposed configuration is plotted in Fig. 1. As it is obvious the bias and baseband signal is driven through the junctions along the *c*-axis of the layered superconductor. We consider the parameters, $\lambda_c$ and $\lambda_{ab}$ to be London penetration depth along the *c*-axis and CuO planes respectively, $\varepsilon_c$ to be the relative permittivity inside the cavity of width $L_x$ respectively, and $\sigma_c$ to be the quasiparticle conductivity across the *c*-axis.

We assume that applying *c*-axis dc bias current, $i_{\text{bias}}$, leads to the terahertz radiation frequency of $\omega_c$, which is considered to be the carrier frequency in our analyses. Now, assuming that the current driven in to the sample has two components of the dc bias and the baseband signal, $i_J = i_{\text{bias}} + i_m$, we investigate the small signal frequency variation due to this current. Here, $i_m$ is related to the data that should be transmitted, $x_m(t)$, by $i_m = \delta x_m(t)$ in which $\delta$ represents the modulation factor. Then, considering that $i_m << i_{\text{bias}}$, the radiation frequency would vary about the central frequency of $\omega_c$, which means that the low frequency signal of $x_m(t)$ is modulated to the terahertz frequency of $\omega_c$.

The Josephson current in the stack of intrinsic Josephson junctions at the presence of the external magnetic field with the rectangular lattice of Josephson vortices at the non-resonant and resonant conditions are obtained from [14]

$$i_{J,\text{rec,non-res}} = \beta\Omega + \frac{\sqrt{\varepsilon_c}\sin(bl/2)}{\eta l b^2 \Omega}, \quad (3)$$

and

$$i_{J,\text{rec,res}} = \beta\Omega + \frac{c\sqrt{\varepsilon_c}\left(1-(-1)^n \cos bl\right)}{\eta l b^2 \Omega^2 s N}, \quad (4)$$

respectively. In the above equations, $l = L_x/\lambda_c$, $\Omega = \omega/\omega_p$, $b = 2\pi\lambda_c dB/\Phi_0$ are cavity width, and resonance frequency, and applied magnetic field, respectively in a reduced system of units, $\beta = 4\pi\sigma_c\lambda_c/\varepsilon_c^{1/2}$ is the dissipation parameter, $\eta = \lambda_{ab}/(s+d)$, $N$ is number of layers, $\Phi_0$ is quantum magnetic flux, and $\omega_p$ is Josephson plasma frequency.

Since we assumed that applying $i_{\text{bias}}$ would result in the radiation frequency of $\omega_c$, then according to the equation 3 and 4, the following relation between $\omega_c$ and $i_{\text{bias}}$ are considered.

$$i_{\text{bias,rec,non-res}} = \beta\omega_c + \frac{\sqrt{\varepsilon_c}\sin(bl/2)}{\eta l b^2 \omega_c}, \quad (5)$$

and

$$i_{\text{bias,rec,res}} = \beta\omega_c + \frac{c\sqrt{\varepsilon_c}\left(1-(-1)^n \cos bl\right)}{\eta l b^2 \omega_c^2 s N}. \quad (6)$$

Expanding equations 3 and 4 about the central carrier frequency, $\omega_c$, leads to the following polynomial series.

$$i_{J,\text{rec,non-res}} = \frac{\sqrt{\varepsilon_c}\sin(bl/2)}{\eta l b^2 \omega_c} + \beta\omega_c + \left[\beta + \frac{\sqrt{\varepsilon_c}\sin(bl/2)}{\eta l b^2 \omega_c^2}\right](\Omega - \omega_c) + O(\Omega - \omega_c)^2 \quad (7)$$

$$i_{J,\text{rec,res}} = \frac{c\sqrt{\varepsilon_c}\left[1-(-1)^n \cos bl\right]}{\eta l b^2 s N \omega_c^2} + \beta\omega_c + \left\{\beta + \frac{2c\sqrt{\varepsilon_c}\left[1-(-1)^n \cos bl\right]}{\eta l b^2 s N}\right\}(\Omega - \omega_c) + O(\Omega - \omega_c)^2 \quad (8)$$

Taking $i_J$ to be $i_{\text{bias}} + i_m(t)$ and $\Omega$ to be $\omega_c + \omega_m(t)$ due to our previous discussion, and subtracting equation 5 and 6 from equations 7 and 8, respectively, one can come to the following expressions for the baseband components of the *c*-axis currents.

$$i_{m,\text{rec,non-res}} = \left[\beta + \frac{\sqrt{\varepsilon_c}\sin(bl/2)}{\eta l b^2 \omega_c^2}\right]\left[\omega_m(t)\right] + O\left[\omega_m(t)\right]^2 \quad (9)$$

$$i_{m,\text{rec,res}} = \left\{\beta + \frac{2c\sqrt{\varepsilon_c}\left[1-(-1)^n \cos bl\right]}{\eta l b^2 s N}\right\}\left[\omega_m(t)\right] + O\left[\omega_m(t)\right]^2 \quad (10)$$

With the assumption of $i_m << i_{\text{bias}}$, one expects that $\omega_m << \omega_c$, which is a normalized parameter in the order of unity. Thus, $\omega_m$ should also be much less than one. This means that we can neglect the quadratic errors in equations 9, and 10. Thus, by substitution of $i_m$ by $\delta x_m(t)$ and neglecting nonlinear term, the following relation between $i_m$ and $\omega_m$ is obtained.

$$\omega_{m,\text{rec,non-res}}(t) \approx \frac{\delta x_{m,\text{rec,non-res}}(t)}{\beta + \frac{\sqrt{\varepsilon_c}\sin(bl/2)}{\eta l b^2 \omega_c^2}} \quad (11)$$

$$\omega_{m,\text{rec,res}}(t) \approx \frac{\delta x_{m,\text{rec,res}}(t)}{\beta + \frac{2c\sqrt{\varepsilon_c}\left[1-(-1)^n \cos bl\right]}{\eta l b^2 s N}} \quad (12)$$

The peak frequency deviation for the corresponding frequency modulation is simply obtained from the comparison of these equations by equation 2.

$$\omega_{\Delta,\text{rec,non-res}} = \frac{\delta}{\beta + \frac{\sqrt{\varepsilon_c}\sin(bl/2)}{l\eta b^2\omega_c^2}} \tag{13}$$

$$\omega_{\Delta,\text{rec,res}} = \frac{\delta}{\beta + \frac{2c\sqrt{\varepsilon_c}\left[1-(-1)^n\cos bl\right]}{l\eta b^2 sN}} \tag{14}$$

Equation (3) for the triangular lattice of Josephson vortices is rewritten as [14]

$$i_{J,\text{tri,res}} \approx \frac{8}{b^4}\alpha_b\Omega + \frac{2\alpha_b\left[\cos(2\Omega l)-\cos(bl)\right]\cos(2\Omega l)}{l\Omega b^2\left[\sin^2(2\Omega l)+(\alpha_{\text{tr}}l)^2\cos^2(2\Omega l)\right]}, \tag{15}$$

with $\alpha_b = \beta+v_{ab}b^2$ and $\alpha_{\text{tr}} = \beta+v_{ab}\Omega^2$, in which $v_{ab}=4\pi\sigma_{ab}/(\gamma^2\varepsilon_c\omega_p)$. The polynomial expansion of the above equation is as follow.

$$\begin{aligned} i_{J,\text{tri,res}} \approx{}& \frac{8}{b^4}\alpha_b\Omega + \frac{2\alpha_b\cos(2\Omega l)\Lambda}{lb^2\omega_c\left[\sin^2(2\Omega l)+(\alpha_{\text{tr}}l)^2\cos^2(2\Omega l)\right]} \\ &+ \left\{\frac{4\alpha_b\cos^2(2\Omega l)\Lambda 2l\left[(\alpha_{\text{tr}}l)^2-1\right]\sin(2\Omega l)}{lb^2\omega_c\left[\sin^2(2\Omega l)+(\alpha_{\text{tr}}l)^2\cos^2(2\Omega l)\right]^2}\right. \\ &- \frac{4\alpha_b l\omega_c\cos(2\Omega l)\sin(2\Omega l)}{lb^2\omega_c^2\left[\sin^2(2\Omega l)+(\alpha_{\text{tr}}l)^2\cos^2(2\Omega l)\right]} \\ &\left.- \frac{2\alpha_b\Lambda\left[\cos(2\Omega l)+2l\omega_c\sin(2\Omega l)\right]}{lb^2\omega_c^2\left[\sin^2(2\Omega l)+(\alpha_{\text{tr}}l)^2\cos^2(2\Omega l)\right]}\right\}(\Omega-\omega_c) \\ &+ O(\Omega-\omega_c)^2 \end{aligned} \tag{16}$$

in which $\Lambda=\cos(2\Omega l)-\cos(bl)$.

In the absence of the external magnetic field, the current flowing through the junction at the resonance mode is related to the radiation frequency by [15]

$$i_{J,\text{res}} = \beta\Omega + \frac{1}{4}\frac{g_m^2(\beta+\beta_\Gamma)\Omega}{\left(\Omega^2-k_m^2\right)^2+(\beta+\beta_\Gamma)^2\Omega^2}, \tag{17}$$

where $\beta_\Gamma$, and $g_m$ are defined by the following equations.

$$\beta_\Gamma = \frac{2N(s+d)\Omega}{\varepsilon_c L_x} \tag{18}$$

$$g_m = \frac{2}{L_x}\int_0^{L_x}\cos\left(\frac{m\pi}{L_x}\right)g(x)\,dx \tag{19}$$

$$k_\omega = \omega/c \tag{20}$$

where $g(x)$ accounts for the possible non-uniformity or spatial modulation of the Josephson critical current. Again, the polynomial expansion method to equation 17 results in the following series.

$$\begin{aligned} i_{J,\text{res}} ={}& \frac{g_m^2(\beta+\beta_\Gamma)\omega_c}{4\left[\left(\omega_c^2-k_m^2\right)^2+(\beta+\beta_\Gamma)^2\omega_c^2\right]} + \beta\omega_c \\ &+ \left\{\beta + \frac{g_m^2(\beta+\beta_\Gamma)\left[\left(\omega_c^2+k_m^2\right)^2-(\beta+\beta_\Gamma)^2\omega_c^2-4\omega_c^4\right]}{4\left[\left(\omega_c^2-k_m^2\right)^2+(\beta+\beta_\Gamma)^2\omega_c^2\right]^2}\right\}\times \\ &(\Omega-\omega_c)+O(\Omega-\omega_c)^2. \end{aligned} \tag{21}$$

On the other hand, the McCumber state which is the model for the non-coupled junctions is chosen as a non-resonating phase state. In this model, the radiating *I-V* characteristics is given by

$$i_{J,\text{non-res}} = \beta\Omega + \frac{\beta}{2\Omega\left(\Omega^2+\beta^2\right)} + \frac{\cos\theta}{\Omega\left(\Omega^2+\beta^2\right)zL}, \tag{22}$$

which can be expanded as below.

$$\begin{aligned} i_{J,\text{non-res}} ={}& \frac{1}{\omega_c\left(\beta^2+\omega_c^2\right)}\left(\frac{\beta}{2}+\frac{\cos\theta}{zL}\right)+\beta\omega_c \\ &+\left[\beta-\left(\frac{\beta}{2}+\frac{\cos\theta}{zL}\right)\frac{\beta^2+3\omega_c^2}{\omega_c^2\left(\omega_c^2+\beta^2\right)^2}\right](\Omega-\omega_c) \\ &+O(\Omega-\omega_c)^2 \end{aligned} \tag{23}$$

In the above equations, $Z=ze^{j\theta}$ is the impedance of the outer space.

Similar to the method presented in details for the case of existing external magnetic field with the rectangular lattice of the Josephson vortices, the peak frequency deviation can be obtained from equations 16, 21, and 23 corresponding to their mentioned conditions of valiance.

$$\begin{aligned} \omega_{\Delta,\text{res}} \approx \delta\Bigg/ & \left\{\frac{4\alpha_b\cos^2(2\Omega l)\Lambda 2l\left[(\alpha_{\text{tr}}l)^2-1\right]\sin(2\Omega l)}{lb^2\omega_c\left[\sin^2(2\Omega l)+(\alpha_{\text{tr}}l)^2\cos^2(2\Omega l)\right]^2}\right. \\ &- \frac{4\alpha_b l\omega_c\cos(2\Omega l)\sin(2\Omega l)}{lb^2\omega_c^2\left[\sin^2(2\Omega l)+(\alpha_{\text{tr}}l)^2\cos^2(2\Omega l)\right]} \\ &\left.- \frac{2\alpha_b\Lambda\left[\cos(2\Omega l)+2l\omega_c\sin(2\Omega l)\right]}{lb^2\omega_c^2\left[\sin^2(2\Omega l)+(\alpha_{\text{tr}}l)^2\cos^2(2\Omega l)\right]}\right\} \end{aligned} \tag{24}$$

$$\omega_{\Delta,\text{res}} = \frac{\delta}{\beta + \frac{g_m^2(\beta+\beta_\Gamma)\left[\left(\omega_c^2+k_m^2\right)^2-(\beta+\beta_\Gamma)^2\omega_c^2-4\omega_c^4\right]}{4\left[\left(\omega_c^2-k_m^2\right)^2+(\beta+\beta_\Gamma)^2\omega_c^2\right]^2}} \tag{25}$$

$$\omega_{\Delta,\text{non-res}} = \frac{\delta}{\beta-\left(\frac{\beta}{2}+\frac{\cos\theta}{zL}\right)\frac{\beta^2+3\omega_c^2}{\omega_c^2\left(\omega_c^2+\beta^2\right)^2}} \tag{26}$$

In either of the considered situations, at the presence and absence of the external magnetic field, the *I-V* characteristics of the resonance and off-resonance radiation are applied. The case of resonance is an appropriate choice for the narrow-band gap signals with the advantage of high radiation power due to the synchronization of the all junctions. Nevertheless, the case of non-resonance radiation has the advantage of modulating wide-band signals though that the obtained power is not very high compared to that of the resonance radiation.

In the non-resonance based modulators, the maximum bandwidth is achieved by adjusting the carrier frequency exactly at the middle of two successive cavity mode frequencies. Thus, this condition results in the maximum bandwidth of equal to the half of the cavity fundamental frequency. Since we have $f=c/2L_x$ for the frequency of the fundamental cavity mode, the theoretical maximum achievable bandwidth would be in the order of $W_{\max}\approx c/4L_x$. According to this relation, small sample sizes are more appropriate for wideband modulators.

The signal to noise ratio, $S/N$, for frequency modulators can be obtained from the following equation [19]:

$$\left(\frac{S}{N}\right)=3\left(\frac{f_\Delta}{W}\right)^2 S_x\gamma \tag{27}$$

In the above equation, $S_x$ is the power of the input signal, and $\gamma$ is its signal to noise ratio. When the modulator is biased in the non-resonance carrier frequency, using the maximum bandwidth, the signal to noise ratio would be at least

$$\left(\frac{S}{N}\right)=3\left(\frac{4L_x f_\Delta}{c}\right)^2 S_x\gamma. \tag{28}$$

Thus, the sample width quadratically controls the signal to noise ratio.

In our numerical analyses, we use the coupled sine-Gordon equations to model the electromagnetic dynamics of the synchronized junctions in the presence and absence of the external magnetic field. We start from the definition of gauge-invariant phase difference as [20-32]

$$\psi_{k+1,k}(\mathbf{r},t)=\phi_{k+1}(\mathbf{r},t)-\phi_k(\mathbf{r},t)-\frac{2\pi}{\Phi_0}\int_{z_k}^{z_{k+1}}A_z(\mathbf{r},t)\,dz, \tag{29}$$

with the phase of order parameter in the $k$-th superconducting layer being $\phi_k$, where $A_z$ stands for the vector potential, the quantum flux is expressed by $\Phi_0$, and $z_k$ is the height of the $k$-th superconducting layer. In this paper, the subscript "$k+1,k$" denotes quantities in the insulating layer between the $k$-th and ($k$+1)th superconducting planes. We assumed that the thicknesses of the superconducting and insulating layers are $s$ and $d$ respectively.

The current through the junctions corresponds to the total pair and quasi-particle tunneling. According to the Josephson relation, the total current density that passes through each junction along the $z$-axis can be obtained from

$$J_{k,k+1}=J_c\sin\varphi_{k+1,k}+\sigma_c E_{k+1,k}+\frac{\varepsilon}{4\pi}\frac{\partial}{\partial t}E_{k+1,k}, \tag{30}$$

where $J_c$ is the Josephson critical current density, and $E_{l+1,l}$ is the $z$-axis component of the electric filed between ($k$+1)th and $k$-th layers. The $c$-axis conductivity, $\sigma_c$, is temperature dependent, and the damping term can be neglected in small values of $\sigma_c$. In the above equation, the first, second, and third terms represent the pair tunneling, quasi-particle tunneling, and the charge accumulation currents respectively. Using the London theory as well as the Maxwell equations, the properties of the system can be expressed as [21-24]

$$\begin{aligned}\frac{\partial^2\varphi_{k+1,k}}{\partial\chi^2}&=\left(1-\zeta\Delta_k^2\right)\left(\frac{\partial\xi_{k+1,k}}{\partial\tau}+\beta\xi_{k+1,k}\right.\\&\qquad\left.+\sin\varphi_{k+1,k}-j_{\text{bias}}-j_m\right)\\\frac{\partial\varphi_{k+1,k}}{\partial\tau}&=\left(1-\alpha\Delta_k^2\right)\xi_{k+1,k},\end{aligned} \tag{31}$$

in which the finite difference operator is defined as $\Delta_n^2 f_n = f_{n+1}+f_{n-1}-2f_n$, the parameters $\alpha=\varepsilon_c\mu^2/sd$ and $\zeta=\lambda_{ab}^2/sd$ are capacitive and inductive couplings between different junctions respectively, and $\mu$ is Thomas-Fermi screening length. The driven current, $j$, in this equation is the sum of the dc bias, $j_{\text{bias}}$, and the baseband signal, $jm$. We have replaced the quantities $\psi_{k+1,k}$, $E_{k+1,k}$, $x$, and $t$, respectively by their equivalents $\varphi_{k+1,k}$, $\xi_{k+1,k}$, $\chi$, and $\tau$ in an scaled unit which is defined by following transformations,

$$\chi=\frac{x}{\lambda_c},\tau=\omega_p t,\ j=\frac{J}{J_c},\xi_{k+1,k}=\frac{\sigma_c}{\beta J_c}E_{k+1,k}, \tag{32}$$

where $J_c$ and $\omega_p$ are the critical current and plasma frequency of the intrinsic junctions that can be replaced by $c\Phi_0/8\pi^2\lambda_c^2 d$ and $c/\varepsilon_c^{1/2}\lambda_c$ respectively. In such a reduced unit, the scaled magnetic field, $b_{l+1,l}$, is also obtained from

$$\frac{\partial\varphi_{k+1,k}}{\partial\chi}=\left(1-\zeta\Delta_k^2\right)b_{k+1,k}, \tag{33}$$

while it is related to the magnetic field through $b_{k+1,k}=2\pi\lambda_c dB_{k+1,k}/\Phi_0$. In the existence of the external magnetic field, the dynamical boundary condition [29] between the oscillatory parts of scaled boundary fields is applied on the edges with the assumption of infinite number of junctions, following Lin *et al.* [24,33] leading to

$$\begin{aligned}\frac{\partial\varphi_{k+1,k}}{\partial\chi}&=\left\langle b_{k+1,k}\right\rangle_\tau+\tilde{b}_{k+1,k}\\\frac{\partial\varphi_{k+1,k}}{\partial\tau}&=j/\beta+\tilde{\xi}_{k+1,k}\\\tilde{b}_{k+1,k}&=\pm z\sqrt{\varepsilon_c/\varepsilon_d}\,\tilde{\xi}_{k+1,k},\end{aligned} \tag{34}$$

where $z$ models the impedance mismatch between the cavity and the dielectric, $\varepsilon_d$, is the relative permittivity of the dielectric, the symbol $<\ldots>_\tau$ stands for time averaging, and the symbol ˜ over field quantities represents their oscillatory parts. The final equation approximates the boundary condition proposed in ref. [18] in the special case of infinite number of junctions where the non-uniformity of oscillations can be neglected. Such a condition is almost satisfied when a large number of junctions are considered in the simulation.

In the case of Josephson oscillation, assuming only parallel to $ab$-axis radiation, the boundary condition is obtained to be [13]

$$\frac{\partial \varphi_{k+1,k}}{\partial \chi} = \left(1 - \zeta \Delta_k^2\right) \tilde{b}_{k+1,k}$$

$$\frac{\partial \varphi_{k+1,k}}{\partial \tau} = \left(1 - \zeta \Delta_k^2\right)\left(j/\beta + \tilde{\tilde{\xi}}_{k+1,k}\right) \tag{35}$$

$$\tilde{b}_{k+1,k} = \pm z \sqrt{\varepsilon_c/\varepsilon_d}\, \tilde{\tilde{\xi}}_{k+1,k},$$

while more precise relations for the electric and magnetic fields are applied.

The initial configuration for the electric field, $E_{k+1,k}$, is set to be its final dc value while the zero initial condition is applied for the gauge-invariant phase difference. These initial values result in the faster convergence as well as preventing from being trapped by the local minima.

## III. Results and Analyses

Equation (31) is solved for $N = 150$ superconducting layers. The values of $\lambda_c$=200µm, $\omega_p$=4.7×10$^{11}$rad/s, $s$=3Å, $d$=12Å, $\alpha$=0.1, $\beta$=0.02, $\zeta$=5×10$^5$, and $\varepsilon_c$=10 are chosen for the parameters. In the numerical solutions, the relative tolerance of the results between successive time steps is set to be lower than 10$^{-7}$ and the mesh size of 0.001 is selected. The more details of the computer simulations, the effects of the bias and geometries on the total characteristics of the mesa structures and the comparison of the theoretically obtained radiation powers with previously published experimental works can be obtained in Ref [34].

### A. Modulator Signal

Fig. 2 presents the typical current-voltage and intensity characteristics for such structures without external magnetic field [34]. As it is obvious in this figure, the intensity is enhanced largely at the resonance modes, which is desirable characteristic for implementing high power modulators. Nevertheless, there exists an intensive dependency of intensity to the radiation frequency near these modes, which widely limits the obtainable bandwidth. On the other hand, though that the radiation power is decreased between resonance modes, it shows an acceptable stability in a wide range of frequency variations, which makes these operation modes appropriate choices for designing wideband modulators.

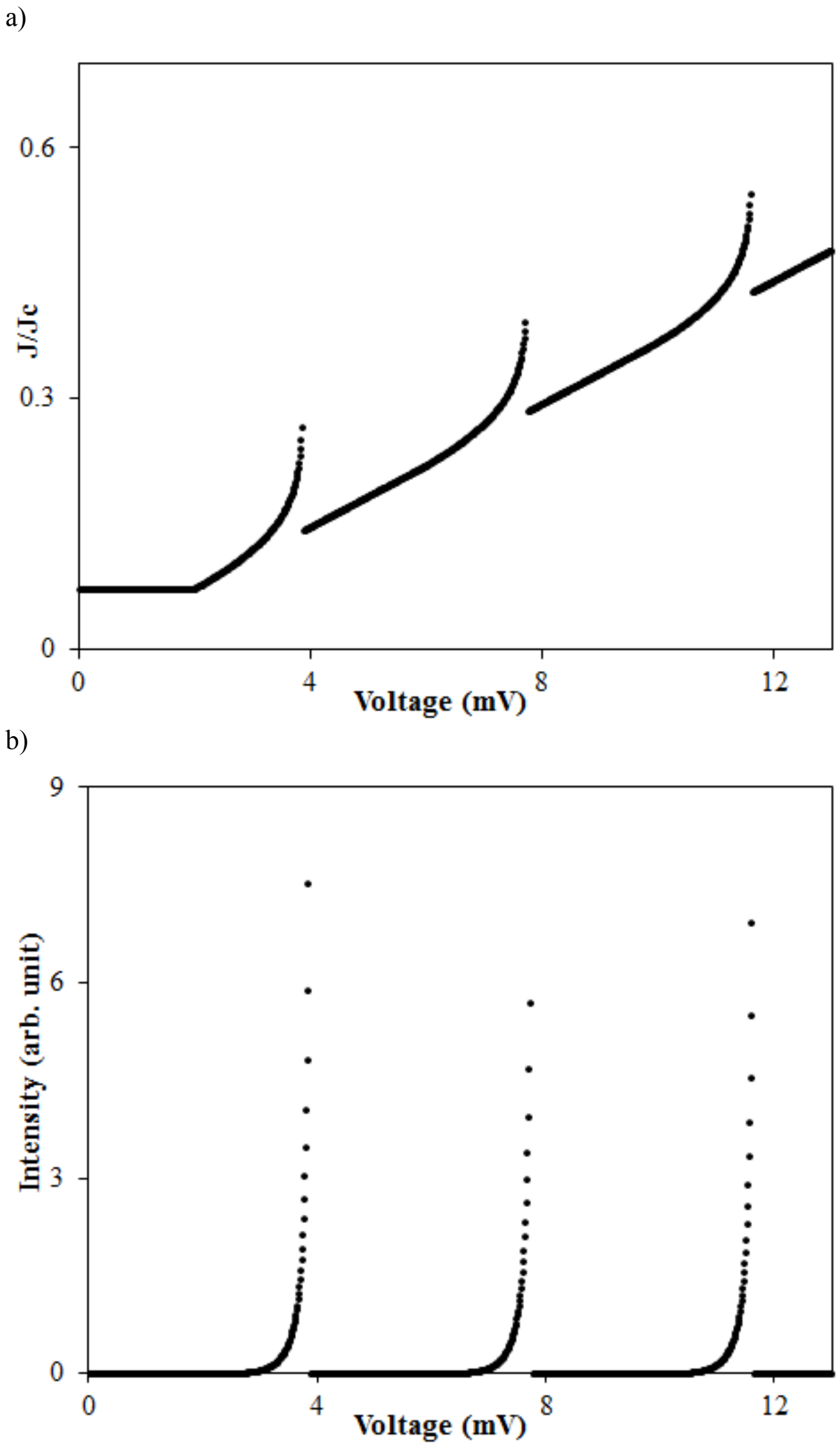

Fig. 2. a) The current-voltage characteristics and b) the radiation power characteristics at the absence of the external magnetic field.

The variation of the peak frequency deviation at different carrier frequencies for the considered conditions is plotted in Fig. 3. The figure shows that the peak frequency deviation obtained for the resonance mode of the rectangular Josephson vortex lattice is independent of carrier frequency. This would be desired characteristic that allows changing the central frequency with the fixed modulation parameter. Consequently, the carrier frequency is adjustable while the phase locking characteristic is not considerably changed. Thus, it could be an appropriate choice for tunable carrier frequency modulators.

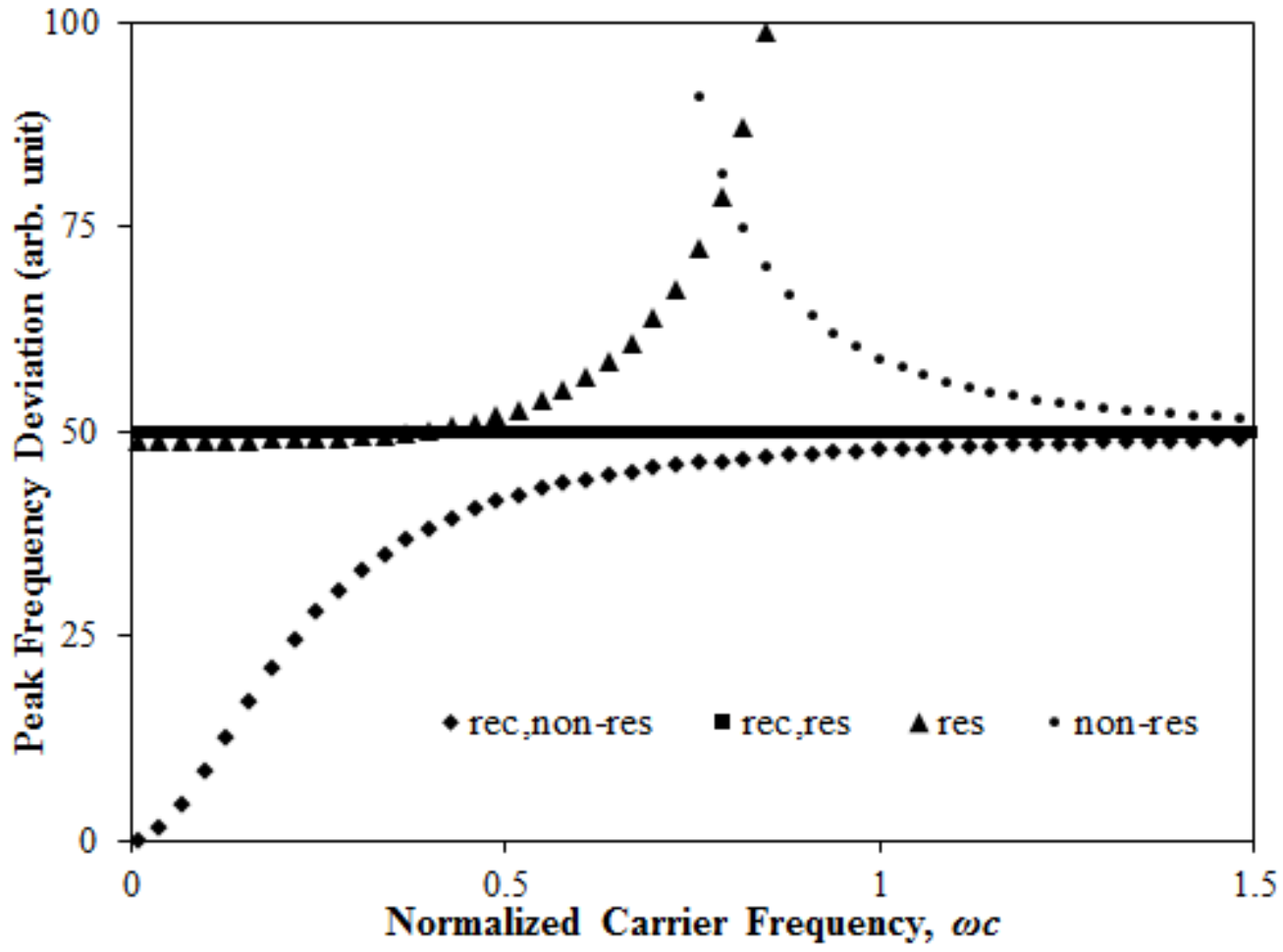

Fig. 3. The dependency of the peak frequency deviation to the normalized carrier frequency.

Figure 4.a presents the numerical results of the modulated output signal after the filtration of its large magnitude low frequency part which is related to the *c*-axis dc conductivity. To obtain this result, in the numerical simulations, the current in the equation 31 is set to be a constant dc value ($J/J_c$=0.2) along with a 50GHz sinusoidal term. Thus, the input data is assumed to be a single harmonic for simplicity. Here, the regime of non-resonating radiation is applied. Since we have $J \approx \sigma_c E$ between two successive cavity resonance modes, the time-varying part of the current, $x_m(t)$, which is the baseband low frequency data, is also reflected on the internal electric field and consequently on the output signal. However, this low frequency signal can be easily filtered and excluded from the output radiation due to the great difference between the frequency ranges. It should be noted that in this figure, slow and fast varying regions represent the baseband signal troughs and peaks, respectively.

In parts b of this figure, the obtained power spectrum corresponding to the output signal of part a is drawn. Using such a plot, the corresponding frequency bandwidth can be estimated. By simultaneous consideration of Figures 4.a and 4.b, it is concluded that at this modulation regime, the bandwidth of about 0.15THz is achieved.

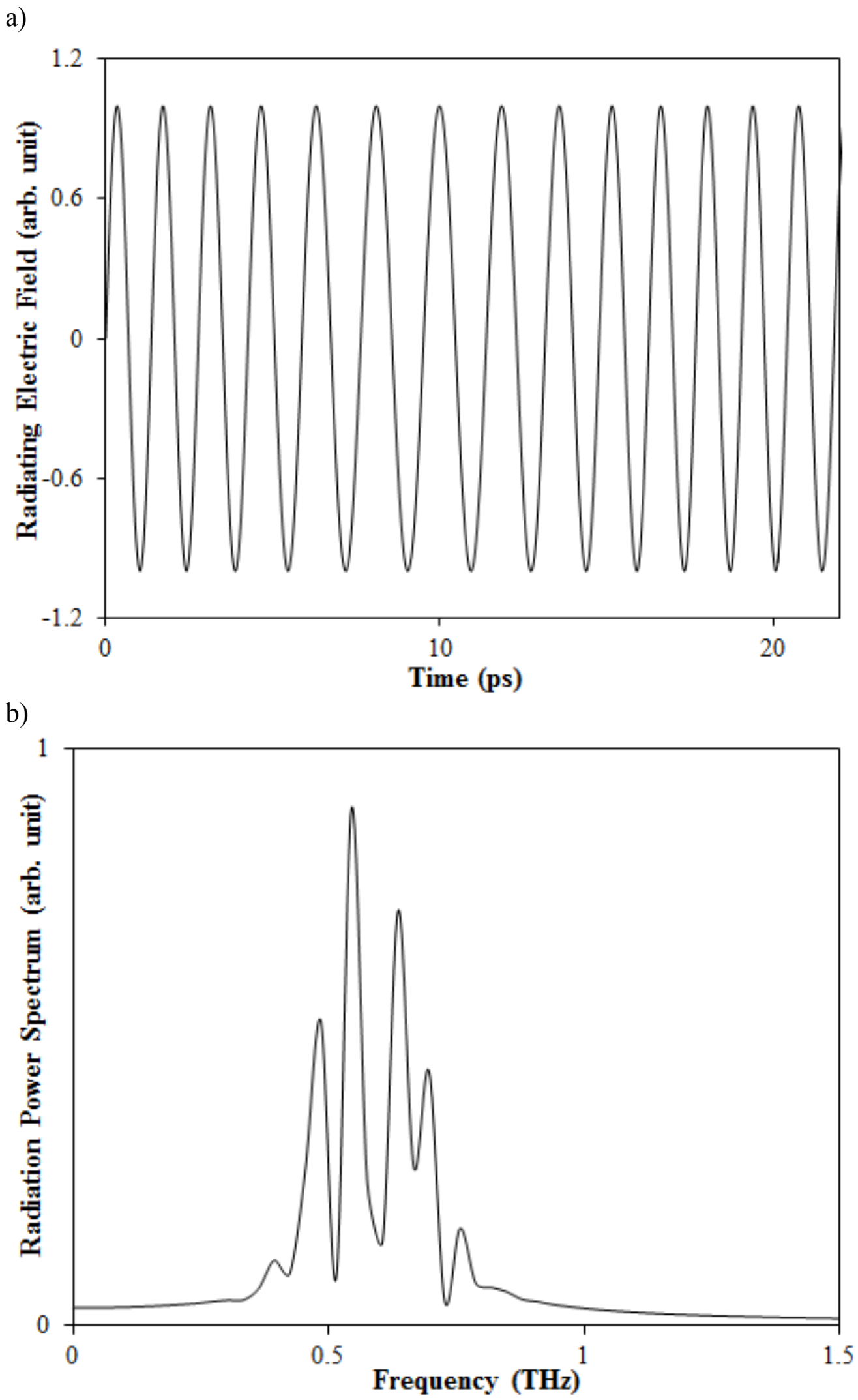

Fig. 4. a) The radiating electric field and b) the power spectrum of the output signal at the non-resonant conditions in the absence of the external magnetic field and the cavity width of 375µm.

## *B. Modulator Characterizations*

Finally, we compared two different regimes of the operation for this kind of modulators, resonant and non-resonant mode. As previously mentioned in this paper, the former has high radiation power and the latter has high frequency bandwidth. These two characteristics are investigated in figure 5. The obtained peak power for the resonant mode is about one order of magnitude larger than that of non-resonant mode, while its bandwidth is one order of magnitude less than the other ones. These results are obtained similar to that of figure 4.

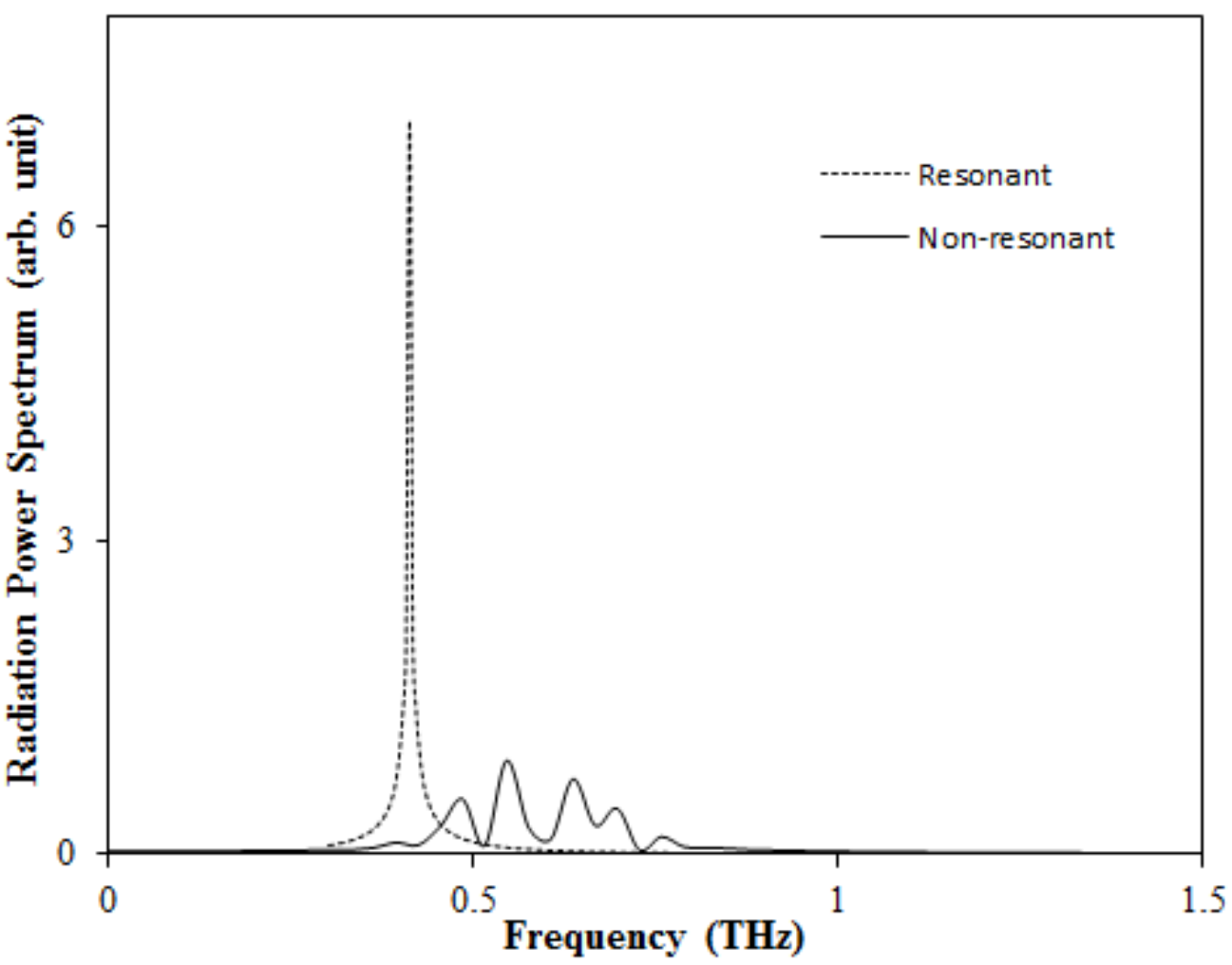

Fig. 5. The obtained power spectrum for two different case of resonant and non-resonant bias

## IV. Conclusion

Applying the stacked Josephson junctions, which is in the molecular structure of the layered high-$T_c$ superconductors as the wideband frequency modulators in the range of terahertz was shown in this paper. We numerically and analytically investigated the characteristic properties of such a current driven modulator. The results show the capability of the rectangular Josephson vortex lattice to be an extremely wideband tunable modulator. In addition, two separate cases of resonant and non-resonant modulators are investigated and compared in this work. The large bandwidth in the order of 0.15THz is achieved for the non-resonant modulator while the radiation power for the resonant one is about one order of magnitude better than that of non-resonant modulator. The carrier frequency can also be adjusted in a wide range from subterahertz to multiterahertz by the electrical bias.